\documentclass[letterpaper,preprintnumbers,prd,twocolumn,nofootinbib,nobibnotes,showpacs]{revtex4}
\usepackage{amsfonts}
\usepackage{mathrsfs}
\usepackage{epsfig}
\usepackage{graphicx}%
\usepackage{dcolumn}
\usepackage{amsmath}

\makeatletter
\def\btt#1{\texttt{\@backslashchar#1}}%
\DeclareRobustCommand\bblash{\btt{\@backslashchar}}%
\makeatother
\begin{document}

\title{A model of nonsingular universe}
\author{Changjun Gao}
\affiliation{$^{1}$The National Astronomical Observatories,
Chinese Academy of Sciences, Beijing 100012}
\affiliation{{$^{2}$Kavli Institute for Theoretical Physics China,
CAS, Beijing 100190, China }}

\date{\today}
\begin{abstract}
In the background of Friedmann-Robertson-Walker Universe, there
exists Hawking radiation which comes from the cosmic apparent
horizon due to quantum effect. Although the Hawking radiation on
the late time evolution of the universe could be safely neglected
, it plays an important role in the very early stage of the
universe. In view of this point, we identify the temperature in
the scalar field potential with the Hawking temperature of cosmic
apparent horizon. Then we find a nonsingular universe sourced by
the temperature-dependant scalar field. We find that the universe
could be created from a de Sitter phase which has the Planck
energy density. Thus the Big-Bang singularity is avoided.
\end{abstract}

\pacs{98.80.Cq, 98.65.Dx}

\maketitle

\section{Introduction}
The standard model of cosmology claims that the primordial
Universe was in a hot, dense, and highly curved state which is
very close to the Big-Bang singularity \cite{wein:72}. The cosmic
energy density, the temperature and other physical quantities
become infinite at the singularity. Thus the presence of infinity
of physical quantities indicates that the Einstein's General
Relativity breaks down at the singularity. General Relativity must
be modified at high densities. In other words, some other aspects
of physics should be taken into account in the gravity theory.

It is usually believed that the Big-Bang singularity might be
regularized in the quantum version of gravity theory. Actually,
the absence of the Big-Bang singularity in a quantum setting could
be expected on qualitative grounds although a completely
self-consistent quantum theory of gravity is still not yet
available so far. As is well known, there are only four
fundamental constants in the theories describing space, time,
matter and thermodynamics. They are the Newton's constant $G$
(related to General Relativity), the speed of light $c$ (related
to Special Relativity), the Planck's constant $\hbar$ (related to
quantum mechanics) and the Boltzmann's constant $k$ (related to
thermodynamics). Consequently, one could have only four quantities
that can be constructed from these four fundamental constants.
They are Planck time $t_p$, Planck length $l_p$, Planck density
$\rho_p$ and Planck temperature $T_p$ \cite{wald:84},
\begin{eqnarray}\label{eq:Planck}
t_{p}&=&\sqrt{\frac{\hbar G}{c^5}}=5.4\times 10^{-44}\ \textrm{s}\;\nonumber\\
l_{p}&=&\sqrt{\frac{\hbar G}{c^3}}=1.6\times 10^{-35}\ \textrm{m}\;\nonumber\\
\rho_{p}&=&\frac{c^5}{\hbar G^2}=5.2\times 10^{96}\ {\textrm{kg}\cdot\textrm{m}^{-3}}\;\nonumber\\
T_{p}&=&\frac{l_p c^4}{Gk}=1.4\times 10^{32}\ \textrm{K}\;
\end{eqnarray}
These quantities are expected to play important role in the
quantum theory of gravity. It is generally assumed that they set
the scale for the quantum gravity effects. The time, length,
energy density and temperature beyond these scales are highly
impossible. In view of this point, the quantum effect should be
considered in General Relativity and the Universe should not be
singular.

The first attempt of taking quantum effect into account in gravity
theory is the elegant work of Hawking \cite{haw:75}. According to
General Relativity, the temperature of a black hole is absolutely
zero. This violates the third law of thermodynamics which states
that one can never approaches the temperature of zero with finite
operations. However, Hawking found that a black hole behaves like
a black body, emitting thermal radiation, with a temperature
inversely proportional to its mass $M$,
\begin{eqnarray}\label{eq:Haw}
T_H=\frac{1}{8\pi M}\cdot\frac{\hbar c^3}{Gk}\;
\end{eqnarray}
In the formulae, four theories of physics, Special Relativity
($c$), General Relativity ($G$), quantum mechanics ($\hbar$) and
thermodynamics ($k$) are present. Then the third law of
thermodynamics is not violated. Motivated by this point, one
expect the Big-Bang singularity may be eliminated by taking into
account the quantum effect. On the other hand, in the background
of Friedmann--Robertson--Walker (FRW) Universe, it is also found
that the universe is filled with the Hawking radiation which comes
from the apparent horizon due to quantum effect. The temperature
of Hawking radiation is give by \cite{cait1,cait2}
\begin{eqnarray}\label{Hawking}
 T_{H}=\frac{1}{2\pi r_A}\cdot\frac{\hbar c}{k}\;
 \end{eqnarray}
where $r_A$ is the radius of cosmic apparent horizon. This
temperature could be observed by the comoving observers in the
universe. We see that not only Special relativity and
thermodynamics but also quantum mechanics are present in the
formulae.

The purpose of this paper is to investigate whether the Big-Bang
singularity could be smoothed by simply taking account of the
cosmic Hawking radiation. The answer is yes. We find that the
universe could be created from a de Sitter phase which has the
Planck energy density. Thus the Big-Bang singularity is avoided.

Actually, many solutions of nonsingular universe have been
presented in the literature. These solutions are based on various
approaches to quantum gravity such as modified gravity models
\cite{non2,non3,non4,non5}, Lagrangian multiplier gravity actions
(see e.g., \cite{non10,non11}), non-relativistic gravitational
actions \cite{non12,non13}, brane world scenarios
\cite{non14,non15} an so on. Here we shall not produce an
exhaustive list of references, but we prefer the readers to read
the nice review paper by Novello and Bergliaffa \cite{nov} and the
references~therein.

The paper is organized as follows. In Section 2, we shall derive
the Friedmann equation and the acceleration equation sourced by
the temperature-dependant scalar field. In Section 3, we present
a nonsingular universe by simply considering the scalar potential
for the oscillators. Section 4 gives the conclusion and
discussion. In the next, we shall use the system of units in which
$G=c=\hbar=k=1$ and the metric signature $(-,\ +,\ +,\ +)$ until
the end of the paper.

\section{Temperature-Dependant Scalar Field}

We consider the Lagrangian density as follows
 \begin{eqnarray}\label{eq:LLL}
{L}=\frac{1}{2}\nabla_{\mu}\phi\nabla^{\mu}\phi+V\left(\phi,\
T\right)\;
\end{eqnarray}
where $V\left(\phi,\ T\right)$ is the temperature-dependant scalar
potential. The theory has been widely studied in the phase
transitions of very early universe \cite{linde:05}. In the
background of Friedmann--Robertson--Walker (FRW) Universe, $T$ could
be taken as the temperature of comic microwave background (CMB)
radiation. The evolution of the temperature is given by
\begin{eqnarray}\label{eq:T}
T_{CMB}\propto\frac{1}{a}\;
\end{eqnarray}
where $a$ is the scale factor of the universe. The observations of
CMB gives the present-day cosmic temperature \cite{komatsu:2009}
\begin{eqnarray}
T_{CMB}\simeq 2.73\ \textrm{K}\;
\end{eqnarray}

On the other hand, in the background of the spatially-flat FRW
Universe, it is found that the universe is filled with the Hawking
radiation which comes from the apparent horizon due to quantum
effect. The temperature of Hawking radiation is give by
\cite{cait1,cait2}
\begin{eqnarray}\label{Hawking}
 T_{H}=\frac{H}{2\pi}\;
 \end{eqnarray}
where $H$ is the Hubble parameter. This temperature could be
observed by the comoving observers in the universe. For the
present-day universe, it is approximately
\begin{eqnarray}
 T_{H}\simeq 10^{-29}\ \textrm{K}\;
 \end{eqnarray}
It is very much smaller than the temperature of CMB. So it is very
safe to neglect the quantum effect for the present-day Universe.
Using the Friedmann equation, the temperature of CMB at the
radiation dominated epoch can be rewritten as
\begin{eqnarray}
T_{CMB}\propto\sqrt{H}\;
\end{eqnarray}
Thus, with the increasing of redshifts, the Hawking temperature
$T_{H}$ increases faster than the CMB temperature $T_{CMB}$. In
other words, the quantum effect would become significant in the
very early universe. In particular, the Hawking temperature at the
Planck time may be such high that
\begin{eqnarray}
 T_{H}\simeq 10^{32}\ \textrm{K}\;
 \end{eqnarray}

Since the Hawking temperature becomes very important in the very
early universe, we identify the temperature in the Lagrangian,
Equation~(\ref{eq:LLL}), with not the CMB temperature but the Hawking
temperature.

Taking into account gravity, we have the action as follows
\begin{eqnarray}
S=\int
d^4x\sqrt{-g}\left[\frac{R}{16\pi}+\frac{1}{2}\nabla_{\mu}\phi\nabla^{\mu}\phi+V\left(\phi,\
T\right)\right]\;
\end{eqnarray}
where $R$ is the Ricci scalar. The metric for the spatially flat
FRW Universe is given by
\begin{eqnarray}
ds^2=-dt^2+a\left(t\right)^2\left(dr^2+r^2d\theta^2+r^2\sin^2\theta
d\varphi^2\right)\;
\end{eqnarray}
where $a(t)$ is the scale factor of the Universe. Then the action
becomes
\begin{eqnarray}
S&=&\int 4\pi r^2dr\int
dt\left[\frac{1}{16\pi}\left(-6\ddot{a}a^2-6\dot{a}^2a\right)\right.\nonumber
\\
&&\left.-\frac{1}{2}a^3\dot{\phi}^2+a^3V\left(\phi,\
H\right)\right]\;
\end{eqnarray}
where dot denotes the derivative with respect to cosmic time $t$.
Here we have replaced the Hawking temperature with the Hubble
parameter in the scalar potential. Variation of the action with
respect to the scale factor $a$ gives the acceleration equation
\begin{eqnarray}\label{eq:acc}
2\dot{H}+3H^2=-8\pi\left[\frac{1}{2}\dot{\phi}^2-V+H\frac{\partial
V}{\partial H}+\frac{1}{3}\left(\frac{\partial V}{\partial
H}\right)^{\cdot}\right]\;
\end{eqnarray}

On the other hand, variation of the action with respect to the
scalar field $\phi$ gives the equation of motion for $\phi$
\begin{eqnarray}\label{eq:eom}
\ddot{\phi}+3H\dot{\phi}+\frac{\partial V}{\partial \phi}=0\;
\end{eqnarray}
Equation~(\ref{eq:acc}) tells us the scalar field contributes a
pressure as the following
\begin{eqnarray}\label{eq:p}
p=\frac{1}{2}\dot{\phi}^2-V+H\frac{\partial V}{\partial
H}+\frac{1}{3}\left(\frac{\partial V}{\partial
H}\right)^{\cdot}\;
\end{eqnarray}
Compared to the pressure of quintessence,
\begin{eqnarray}\label{eq:quin}
p=\frac{1}{2}\dot{\phi}^2-V\;
\end{eqnarray}
the last two terms on the right hand side of Equation~(\ref{eq:p}) come
from the variation of scalar potential with respect to the Hawking
temperature.

In order to find the energy density $\rho$ contributed by the
scalar field, we put
\begin{eqnarray}\label{eq:rho0}
\rho=\frac{1}{2}\dot{\phi}^2+V+F\;
\end{eqnarray}
where $F$ is a function to be determined. Substituting
Equation~(\ref{eq:p}) and Equation~(\ref{eq:rho0}) into the energy
conservation equation
\begin{eqnarray}\label{eq:ece}
\dot{\rho}+3H\left(\rho+p\right)=0\;
\end{eqnarray}
and taking into account Equation~(\ref{eq:eom}), we obtain
\begin{eqnarray}\label{eq:F}
\left(F+H\frac{\partial V}{\partial
H}\right)^{\cdot}+3H\left(F+H\frac{\partial V}{\partial
H}\right)=0\;
\end{eqnarray}
So we get
\begin{eqnarray}
F=-H\frac{\partial V}{\partial H}+\frac{F_0}{a^3}\;
\end{eqnarray}
$F_0$ is an integration constant. Since there is no constant in
the Lagrangian density, Equation~(\ref{eq:LLL}), we expect $F_0$ should
be zero. So the energy density is given by
\begin{eqnarray}\label{eq:rho}
\rho=\frac{1}{2}\dot{\phi}^2+V-H\frac{\partial V}{\partial H}\;
\end{eqnarray}
The last term comes from the variation of the scalar potential
with respect to the Hawking temperature.

Now we can write the Friedmann equation as the following
\begin{eqnarray}\label{eq:Fri}
3H^2=8\pi\left(\frac{1}{2}\dot{\phi}^2+V-H\frac{\partial
V}{\partial H}\right)\;
\end{eqnarray}
Equations~(\ref{eq:acc}),\ (\ref{eq:eom}), and\ (\ref{eq:Fri}) constitute the
main equations which govern the evolution of the universe. Among
the three equations, only two of them are independent. But we have
three unknown functions, $a(t)$, $\phi(t)$ and $V(\phi,\ T)$. So
we are left with one freedom. For simplicity, we may fix the
expression of the scalar potential.

\section{A Nonsingular Universe}

We consider one of the simplest scalar potentials given by
\begin{eqnarray}\label{eq:Pot}
V\left(\phi,\ T\right)=2\pi^2\eta^2T^2\phi^2\;
\end{eqnarray}
with $\eta$ a positive dimensionless constant. This form of
temperature-dependant potential comes from the high-temperature
expansion of the \emph{finite-temperature effective potential}
\cite{KT:1990}:
\begin{eqnarray}
V_T\left(\phi,\
T\right)=V\left(\phi\right)+\frac{\lambda}{8}T^2\phi^2-\frac{\pi^2}{90}T^4+\cdot\cdot\cdot\;
\end{eqnarray}
where $\lambda$ is a constant. For simplicity and to the first
order of temperature corrections, we are only interested in the
second term in the right hand of the equation.

Using the formulae of temperature, Equation~(\ref{Hawking}), we can
rewrite the scalar potential as follows
\begin{eqnarray}\label{eq:Pot1}
V\left(\phi,\ T\right)=\frac{1}{2}\eta^2H^2\phi^2\;
\end{eqnarray}
Now the Friedmann equation, Equation~(\ref{eq:Fri}), is reduced to
\begin{eqnarray}\label{eq:Fri1}
3=4\pi\left[\left(\frac{d\phi}{dx}\right)^2-\eta^2\phi^2\right]\;
\end{eqnarray}
where $x$ is defined by
\begin{eqnarray}
x\equiv\ln a\;
\end{eqnarray}
 Solving the differential equation, we find
\begin{eqnarray}
\phi=\sqrt{\frac{3}{4\pi}}\frac{a^{\eta}-a^{-\eta}}{2\eta}\;
\end{eqnarray}
Without the loss of the generality, we set the integration
constant to be zero.

Substituting it into the equation of motion, Equation~(\ref{eq:eom}),
and using Equation~(\ref{eq:rho}) we obtain the energy density
\begin{eqnarray}
\rho=\frac{\rho_p a^{4\eta}}{a^6\left(1+a^{2\eta}\right)^4}\;
\end{eqnarray}
where $\rho_p$ is the integration constant. We will shortly find
that it is the Planck energy density. Substituting the energy
density into the energy conservation equation, Equation~(\ref{eq:ece})
we obtain the equation of state~$w$
\begin{eqnarray}
w\equiv\frac{p}{\rho}=1-\frac{4}{3}\eta\cdot\frac{1-a^{2\eta}}{1+a^{2\eta}}\;
\end{eqnarray}
In order that the energy density is not divergent when $a=0$, we
should require that
\begin{eqnarray}
\eta\geq\frac{3}{2}\;
\end{eqnarray}
On the other hand, in order that the equation of state parameter
is not less than minus one when $a=0$, we should require that
\begin{eqnarray}
\eta\leq\frac{3}{2}\;
\end{eqnarray}
So $\eta$ is forced to be
\begin{eqnarray}
\eta=\frac{3}{2}\;
\end{eqnarray}
Then the energy density is given by
\begin{eqnarray}
\rho=\frac{\rho_p}{\left(1+a^{3}\right)^4}\;
\end{eqnarray}
So the energy density increases with the increasing of redshifts.
When $a=0$, we have the maximum energy density
\begin{eqnarray}
\rho={\rho_p}\;
\end{eqnarray}
It is generally believed that the Planck energy density may  be
the maximum density in the universe. So the integration constant
$\rho_p$ is actually the Planck density. The evolution of the
scale factor is found to be
\begin{eqnarray}
\ln a+\frac{1}{6}a^6+\frac{2}{3}a^3=\sqrt{\frac{8\pi}{3}}\cdot
\frac{t}{t_p}\;
\end{eqnarray}
where $t_p$ is the Planck time.

\begin{figure}
\includegraphics[width=8.5cm]{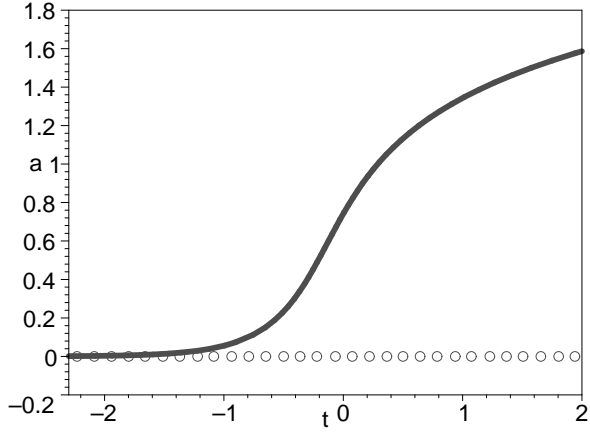}
\\
\caption{The evolution of the scale factor $a$ with respect to the
cosmic time $t$. It shows that the scale factor approaches zero
when $t=-\infty$. The unite of time is the Planck time $t_p$.}
\label{fig:scale}
\end{figure}

In Figure~\ref{fig:scale}, we plot the evolution of the scale factor
with respect to the cosmic time. It shows that the scale factor
approaches zero when $t=-\infty$. The scale factor is actually
physically meaningless. So we do not worry its vanishing at
$t=-\infty$. It is the radius $s$ of Hubble horizon that describes
the physical size of the universe
\begin{eqnarray}
s\equiv\frac{1}{H}\;
\end{eqnarray}

\begin{figure}
\includegraphics[width=8.5cm]{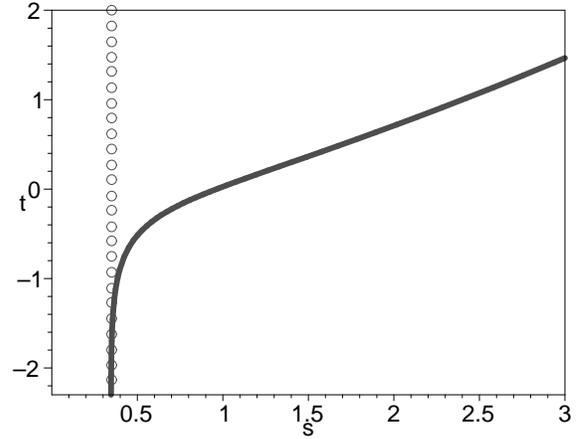}
\\
\caption{The evolution of the Hubble radius $s$ with respect to
the cosmic time $t$. It shows that the Hubble horizon approaches a
finite constant when $t=-\infty$. In other words, the physical
size of the universe is bounded below. The unit of time and length
are the Planck time $t_p$ and Planck length $l_p$, respectively.}
\label{fig:hubble}
\end{figure}
\begin{figure}
\includegraphics[width=8.5cm]{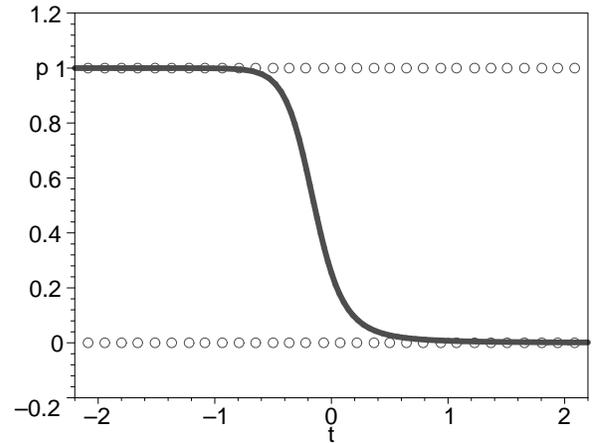}
\\
\caption{The evolution of energy density $\rho$ with respect to
the cosmic time $t$. It shows that the energy density approaches
the Planck density when $t=-\infty$. The unit of time and energy
density are the Planck time $t_p$ and the Planck density $\rho_p$,
respectively.} \label{fig:rho}
\end{figure}

In Figure~\ref{fig:hubble}, we plot the evolution of the Hubble
horizon $s$ with respect to the cosmic time $t$. It shows that the
Hubble horizon approaches a finite constant when $t=-\infty$.
Therefore, the universe is nonsingular. In Figure~\ref{fig:rho}, we
plot the evolution of the energy density $\rho$ with respect to
the cosmic time $t$. It shows that the energy density approaches
the Planck density when $t=-\infty$. Since the universe is created
with the finite Hubble scale and finite energy density, it is a
nonsingular universe. We note that the cosmic time $t$ is real and
physical. So the minus infinity $t=-\infty$ is real and geodesic
one.

\section{Conclusion and Discussion}

Although the Hawking radiation on the late time evolution of the
universe could be safely neglected, it plays an important role in
the very early stage of the universe. In view of this point, we
identify the temperature in the scalar field potential with the
Hawking temperature of cosmic apparent horizon. Then we find a
nonsingular universe sourced by the Hawking-temperature-dependant
scalar field. We had better point out that the analysis of
singularity here needs more attentions. For example, we need to
modify gravity in a non-perturbation fashion to really make sense
\cite{am:2012}.

The extension of the idea to the general scalar-tensor theories is
straightforward:
\begin{eqnarray}\label{eq:LLLL}
{L}=f\left(R,\ \frac{1}{2}\nabla_{\mu}\phi\nabla^{\mu}\phi,\
\phi,\ T\right)\;
\end{eqnarray}
with $f$ an arbitrary function of the variables. But we note that
the whole approach of this paper can be applied only to the FRW
universe, because for a general geometry the Hawking temperature
has to be defined separately. As another example, we may take
$\phi$ as a constant and take the potential as follows
\begin{eqnarray}
V\propto T^4\;
\end{eqnarray}
Then we would have the Friedmann equation as follows
\begin{eqnarray}
3H^2=16\pi\rho_p\left(1-\sqrt{1-\frac{\rho}{\rho_p}}\right)\;
\end{eqnarray}
where $\rho$ may be the energy density of radiation, dark matter
and dark energy. It is apparent the maximum of $\rho$ is the
Planck density. It reveals a nonsingular universe. When
$\rho\ll\rho_p$, it goes back the usual Friedmann equation.

On the other hand, if we take the potential as
\begin{eqnarray}
V\propto \frac{1}{T^2}\;
\end{eqnarray}
we would have the Friedmann equation as follows
\begin{eqnarray}
3H^2=4\pi\rho\left(1+\sqrt{1-\frac{\rho_X^2}{\rho^2}}\right)\;
\end{eqnarray}
where $\rho_X$ is some constant. It is apparent the minimum of
$\rho$ is $\rho_X$. It reveals a nonsingular but future universe.
When $\rho_X\leq\rho$, it restores to the usual Friedmann
equation. Of course, when $T=const$, it is the quintessence.

\section{Acknowledgements}

I thank the anonymous referees for the expert and insightful
comments, which have certainly improved the paper significantly. I
thank Hao Wei and Yunsong Piao for helpful discussions.
This work is supported by the National Science Foundation of China
under the Key Project Grant 10533010, Grant 10575004, Grant
10973014, and the 973 Project (No. 2010CB833004).

\bibliographystyle{mdpi}

\makeatletter

\renewcommand\@biblabel[1]{#1. }

\makeatother

\newcommand\AL[3]{~Astron. Lett.{\bf ~#1}, #2~ (#3)}
\newcommand\AP[3]{~Astropart. Phys.{\bf ~#1}, #2~ (#3)}
\newcommand\AJ[3]{~Astron. J.{\bf ~#1}, #2~(#3)}
\newcommand\APJ[3]{~Astrophys. J.{\bf ~#1}, #2~ (#3)}
\newcommand\APJL[3]{~Astrophys. J. Lett. {\bf ~#1}, L#2~(#3)}
\newcommand\APJS[3]{~Astrophys. J. Suppl. Ser.{\bf ~#1}, #2~(#3)}
\newcommand\JHEP[3]{~JHEP.{\bf ~#1}, #2~(#3)}
\newcommand\JCAP[3]{~JCAP. {\bf ~#1}, #2~ (#3)}
\newcommand\LRR[3]{~Living Rev. Relativity. {\bf ~#1}, #2~ (#3)}
\newcommand\MNRAS[3]{~Mon. Not. R. Astron. Soc.{\bf ~#1}, #2~(#3)}
\newcommand\MNRASL[3]{~Mon. Not. R. Astron. Soc.{\bf ~#1}, L#2~(#3)}
\newcommand\NPB[3]{~Nucl. Phys. B{\bf ~#1}, #2~(#3)}
\newcommand\PLB[3]{~Phys. Lett. B{\bf ~#1}, #2~(#3)}
\newcommand\PRL[3]{~Phys. Rev. Lett.{\bf ~#1}, #2~(#3)}
\newcommand\PR[3]{~Phys. Rep.{\bf ~#1}, #2~(#3)}
\newcommand\PRD[3]{~Phys. Rev. D{\bf ~#1}, #2~(#3)}
\newcommand\RMP[3]{~Rev. Mod. Phys.{\bf ~#1}, #2~(#3)}
\newcommand\SJNP[3]{~Sov. J. Nucl. Phys.{ ~#1}, #2~(#3)}
\newcommand\ZPC[3]{~Z. Phys. C{\bf ~#1}, #2~(#3)}
\newcommand\CQG[3]{~Class. Quant. Grav.{~#1}, #2,~#3}
\newcommand\CMP[3]{~Commun. Math. Phys.{\bf ~#1}, #2~(#3)}

\end{document}